\documentclass[11pt]{article}

\pdfoutput=1

\usepackage[english]{babel}
\usepackage{xcolor} 
\usepackage{graphicx}
\usepackage{hyperref}
\usepackage{booktabs}

\begin{document}

\title{State machine inference of QUIC}

\author{Abdullah Rasool\\Amsterdam, The Netherlands
\and
Greg Alp\'ar\\Open University, Radboud University\\Nijmegen, The Netherlands
\and
Joeri de Ruiter\\SIDN Labs\\Arnhem, The Netherlands
}

\date{}

\maketitle

\begin{abstract}
QUIC is a recent transport protocol that provides reliable, secure and quick service on top of UDP in the internet. As QUIC is implemented in the application space rather than in the operating system's kernel, it is more efficient to dynamically develop and roll out. Currently, there are two parallel specifications, one by Google and one by IETF, and there are a few implementations. In this paper, we show how state machine inference can be applied to automatically extract the state machine corresponding to the protocol from an implementation. In particular, we infer the model of Google's QUIC server. This is done using a black-box technique, making it usable on any implementation of the protocol, regardless of, for example, the programming language the code is written in or the system the QUIC server runs on. This makes it a useful tool for testing and specification purposes, and to make various (future) implementations more easily comparable.
\end{abstract}


\section{Introduction}

Internet applications require reliable, secure and fast communication. Reliability is most often achieved by the Transmission Control Protocol (TCP) and security is provided by the Transport Layer Security (TLS) protocol. Both of these protocols run in the end nodes, i.e. in clients and servers. Although the high speed is fundamentally maintained by the network infrastructure, it has to be preserved for the applications. TCP and TLS, underlying these applications, have a crucial role in the performance perceived by the user.

Quick UDP Internet Connections (QUIC) is Google's recent protocol to offer a more efficient alternative to the reliable TCP and secure TLS. It aims at shorter connection establishment, multiplexed streams and applies several other optimization mechanisms. QUIC transports packets over the User Datagram Protocol (UDP), a widely deployed light-weight transport-layer protocol, and runs within the application as opposed to the kernel. Therefore, QUIC is much easier to incrementally develop and roll out than TCP or UDP.

Developed first in 2013, QUIC has constantly been improved, making its implementation and specification in a state of flux. A protocol specification of a distributed program is important not only for implementation, but also interconnectivity and testing purposes. To capture the essence of a specification, a state machine can define all possible states, including responses to incoming messages in every state. Based on that state machine, implementations can be tested. We note that typically, the client side and the server side implement different states.

Ideally, a state machine could be provided in the specifications of a protocol, though in practice this is often not the case.  Such a model is not yet given for QUIC. It is, however, possible to automatically extract a state machine from an implementation. This method is known as state machine inference or model learning. The resulting state machine can subsequently be used to analyse the implementation.

\paragraph{Contribution} In this paper, we infer a state machine of one of the implementations of Google's QUIC server. Such a model can reveal bugs or contradictions. In QUIC's case, the implementation fortunately turns out to be sound. It merely shows one undocumented transition---which turns out to be harmless. Besides the state machine, we explore a library used for learning models from implementations and develop some tooling for QUIC. This makes it easier to test other QUIC implementations in the future and supports development of the protocol.


\subsection{Related Work}

Previously, state machine inference has been used successfully to extract state machine models from implementations of several other network protocols, including TCP~\cite{fiteruau16-tcp}, TLS~\cite{deruiter15-tls} and Wi-Fi \cite{esorics2018-wifi}.


Kakhki et al.~\cite{Kakhki:2017:TLL:3131365.3131368} construct a state machine of QUIC by looking at the log files created during the execution of the protocol. Their goal is to run QUIC in a large number of different environments (several mobile and desktop devices with different operating systems) and to use the state machine to understand differences across QUIC versions and the different environments. While their approach is passive, in this paper, we perform active learning of the QUIC protocol, observing the output of the implementation when actively sending input to it. Passive learning has a downside in that it depends on the states that previous executions have seen. If past executions have not entered edge cases, then the learned model is limited. This does not hold for active learning, in which we send arbitrary requests in order to reach the edge cases too. Next to this, our approach is completely black-box, that is, we do not require any access to the application we are analysing, apart from a network connection to send and receive QUIC messages.

Google claims that QUIC offers performance improvements~\cite{carlucci2015http}, which is reinforced by the product name that invites for further studies. Megyesi et al.~\cite{megyesi2016quick} show that in more than 40\% of the studied scenarios, the page load times significantly improved with the experimental version of QUIC compared to traditional TCP and HTTP/1.x. 
Another field where QUIC has improvements over existing procotols is security. It provides a secure (authenticated and encrypted) channel by default. In~\cite{lychev2015secure}, Lychev et al.~report a possibility for the handshake to fail due to an inconsistent state between the client and the server. 
A formal analysis of the protocol did show that QUIC's multi-state key exchange meets the security properties as suggested by the designers~\cite{fischlin2014multi}.
Based on the model that is presented in this work, implementations might be simplified and enhanced.

\section{QUIC}



QUIC is a transport protocol designed to improve performance. It is built on top of UDP in the user space, which allows for rapid deployment of changes in QUIC. 

\begin{figure}
	\centering
	\includegraphics[width=.8\linewidth]{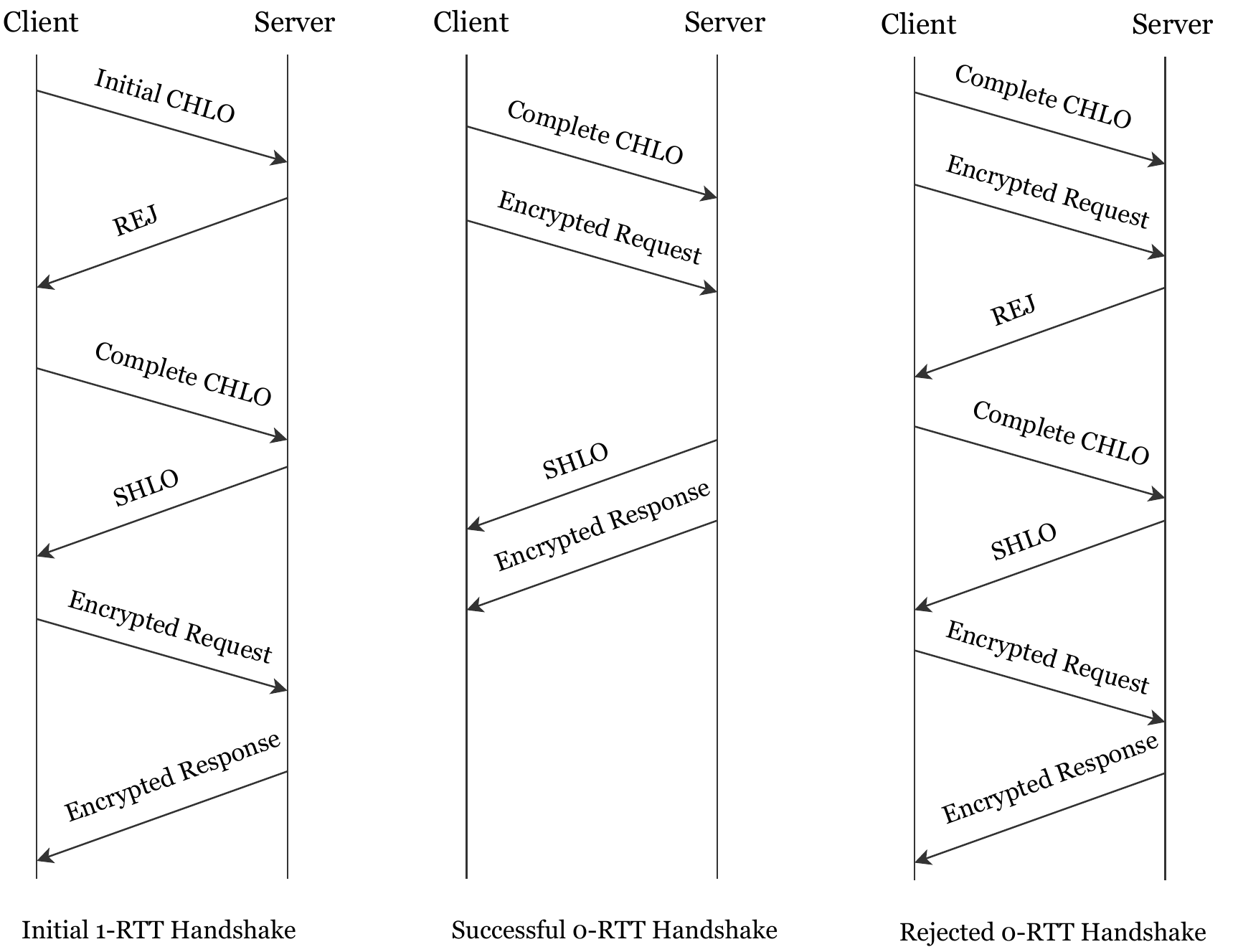}
	\caption{Overview of different QUIC handshakes \cite{langley2017quic}}
	\label{fig:handshakesquic}
\end{figure}

There are two major design decisions that enable QUIC to have an improved performance compared to TCP. First, it combines the cryptographic and transport handshake to reduce the set-up latency and to provide a secure channel by default. To achieve this, QUIC provides three types of connection establishment, see Figure~\ref{fig:handshakesquic}.

\begin{itemize}
	\item \textbf{Initial handshake} (or \texttt{Initital 1-RTT handshake}): The client initially has little to no information about the server. The client starts with a client hello (\texttt{CHLO}) message which will be rejected with a \texttt{REJ} message by the server. This contains a server configuration with its long-term Diffie--Hellman public value, a certificate chain authenticating the server and a timestamp. Now the client can send a new complete \texttt{CHLO} message containing its initial tags and the received ones from the \texttt{REJ} message. If the handshake is successful, the server responds with an encrypted server hello (\texttt{SHLO}) message. The \texttt{SHLO} message includes the server's ephemeral Diffie--Hellman public value, which is used to compute the ephemeral session key.
	\item \textbf{Repeat handshake} (or \texttt{0-RTT handshake}): The client has already seen the \texttt{REJ} message in some previous connection establishment. It stored the tags from the \texttt{REJ} message so that it can craft the complete \texttt{CHLO} message at once. Again, if the handshake is successful, the server responds with an encrypted \texttt{SHLO} message. Using the initial shared key, both parties can compute the ephemeral keys to send and receive any further messages. If the client wishes to achieve 0-RTT latency, then it must encrypt the request with the initial keys and send it before it receives an answer from the server. In order to achieve this, the server also stores the client's nonce and its public value such that it can compute the shared key. 
	\item \textbf{Failed 0-RTT} (or \texttt{Rejected 0-RTT handshake}): If the server information expired in the complete \texttt{CHLO}, the server responds with a \texttt{REJ} message. In this case, the 0-RTT attempt failed, and the handshake continues as if it was an initial handshake.
\end{itemize} 

The second design decision to improve performance addresses the problem of head-of-line blocking. This occurs when a packet is lost in transit and must be retransmitted. TCP's reliable service guarantees that packets are delivered in the same order as they have been sent. This makes TCP traffic susceptible to head-of-line blocking: all subsequent packets must wait until the lost packet has been received. QUIC uses multiple streams---lightweight ordered abstractions---to reduce this problem.

Every stream is cut into frames. Various types of frames exist. First, regular stream frames carry data used for connection establishment (e.g.\ \texttt{CHLO}, \texttt{REJ}). These use the fixed stream 1. Next, acknowledgement frames, discussed later, notify the sender about successful packet delivery. There is also a frame used for congestion control, which we do not use in our setting since we only have a single user on a local network. Lastly, there are two frame types to close the connection; they are discussed later.

If a packet is lost, it only impacts those streams of which data was carried in that packet. Subsequent data received on other streams is continued to be reassembled and delivered to the application \cite{grigorik2013high,langley2017quic}. 


\section{State machine inference}
To learn the state machine that is implemented for a protocol, we can use state machine inference. This technique extracts the model from an implementation in a black-box fashion. The implementation is then called the System Under Learning (SUL). The inference is performed using two types of actions: 1. Sending sequences of messages (queries) to the implementation and observing the corresponding response; and 2. Resetting the SUL to its initial state so that two queries can be executed independently. Note that this black-box approach works on any implementation of the corresponding protocol.

The process to infer a state machine consists of two steps. First, a \emph{learning algorithm} is used to come up with a hypothesis of the implemented state machine. For this we use the L* algorithm, originally published by Angluin \cite{angluin1987learning} and extended by Niese to apply it to Mealy machines \cite{niese2003integrated}. Once this algorithm produces a hypothesis, an \emph{equivalence algorithm} is run. This algorithm verifies whether the produced model matches the implementation. This is done by sending messages to the implementation and checking whether the responses match the hypothesis. In case of a mismatch, the corresponding message--response sequence is provided to the learning algorithm, which updates its hypothesis. Using the updated one, the algorithm continues until finding an acceptable hypothesis. This hypothesis is then provided to the equivalence algorithm again for verification. This process continues until the equivalence algorithm deems the model correct and the state machine is provided. Note again that because of the black-box approach, we can never be sure that the model we found is complete. The equivalence algorithm sends a number of queries to the SUL and if all the responses match the model, it is accepted. The number of queries, as well as the minimum and maximum number of messages per query are parameters to the algorithm.

\begin{figure}
\includegraphics[width=\columnwidth]{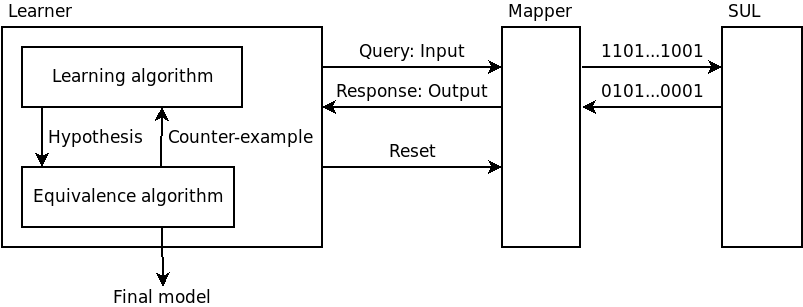}
\caption{State machine inference setup}
\label{fig:inference}
\end{figure}

The learning and equivalence algorithms work with abstract messages, or symbols, that are part of the input and output alphabet. Of course, the SUL does not understand these abstract symbols; so, a component is needed in between the algorithms and the SUL to translate between the abstract symbols and the actual protocol messages. This component is called the \emph{mapper}. It is basically a stateless implementation of the protocol, though it might have to keep track of some minimal state. This might be necessary in order to be able to successfully complete protocol sessions. For example, if the protocol contains some kind of key exchange, the mapper needs to keep track of the information exchanged that is necessary to compute the final key. An overview of the setup for state machine inference can be found in Figure~\ref{fig:inference}. 

\section{Setup}

In this section we describe the concrete setup that we used to infer the state machine of the QUIC server. 

\subsection{Learner}

For the learner we made use of the tool StateLeaner\footnote{\url{https://github.com/jderuiter/statelearner}}. This tool is a wrapper around the LearnLib\footnote{\url{https://github.com/LearnLib/learnlib}} library, which implements several learning and equivalence algorithms. StateLearner allows us to make use of these algorithms with little effort. It only requires a specification of the abstract input alphabet (in the form of a list of strings), after which it can be connected to our QUIC specific mapper via a socket. After setting the desired parameters, StateLearner can be run and will produce the final model.

\subsection{Mapper}


The mapper is tasked with translating the abstract symbols sent by the learner into concrete QUIC messages that can be sent to the SUL and vice versa for the responses from the SUL. The mapper needs to be capable of sending messages in arbitrary order. This makes the mapper basically a stateless QUIC client. 

Initially, we tried to adapt the original QUIC client into such a mapper. Unfortunately, this did not work as it was far from trivial to work with the internals of the original client. We decided to create our own minimal client, which allowed more flexibility for the analysis. The code to craft and interpret the QUIC packets is implemented using Scapy\footnote{\url{https://scapy.net/}}.

We used the limited documentation available at \cite{Britt2016} to be able to understand the packet structure of QUIC. Unfortunately, the information was not very extensive, making it necessary to also use the original QUIC source code as a reference. Importantly, certain frames within streams transmitted in a packet. This made it challenging to create a minimal QUIC client. Table~\ref{tbl:messages} contains the messages that the client can construct. The source code of the mapper is available on GitHub\footnote{\url{https://github.com/aredev/quic-scapy}}.


A challenge to be resolved to infer models of QUIC, is the handling of non-deterministic behavior. This problem arises from the difference in behavior between regular clients and our learning setup. A regular client typically goes through the protocol flow as fast as possible and knows what messages to expect. In our learning setup, however, we send messages in arbitrary order and do not know yet what message to expect in response to a particular message---if any at all. As a result, we have to wait for some time to determine what the response was or to decide whether no response was sent by the server. However, in QUIC there is no link between requests and responses; a retransmission to a previous request may be seen as a response to the current request. To handle this, we use an ad-hoc solution to infer models of QUIC. In order to increase certainty that a particular response belongs to a given request, we repeat the request three times. We then measure the occurrences of the response and the most frequent one is considered to be the actual response. This does not hold in all cases. For example, a connection can only be closed once. These exceptions are handled with manual response filtering by isolating the request and sending it several times. If during model learning we encounter an unseen response, we test it manually and add it to the filter.

\begin{table}\scriptsize
\centering
\renewcommand\arraystretch{1.5}
 \begin{tabular}{p{2cm} p{2cm} p{6cm} }
	\toprule 
	\textbf{Learning symbol} & \textbf{Concrete QUIC Request} & \textbf{Explanation} \\ 
	\midrule 
	INIT-CHLO & Initial CHLO request & Starts a fresh new connection and is used when connecting to a previously unknown server. \\ 
	FULL-CHLO & Complete CHLO Request & Uses the data from the initial \texttt{CHLO} with the missing tags received in a rejection message. This does not create a new connection ID unlike the \texttt{INIT-CHLO} message. Instead, it uses the previous connection ID. If there is no such value, then it defaults to -1. \\ 
	0RTT-CHLO & Complete CHLO request & Starts a fresh connection but uses the stored tags which were missing from a previous initial \texttt{CHLO}.  \\ 
	GET & HTTP/2 GET Request GET (Stream Frame) & Makes an \texttt{HTTP/2 GET} request for the fixed domain \url{www.example.org} \\ 
	CLOSE & Connection Close Frame & Notifies that the connection is being closed. If there are streams in flight, those streams are all implicitly closed when the connection is closed.  \\ 
	\bottomrule 
 \end{tabular}    
\caption{QUIC messages supported by the mapper}
\label{tbl:messages}
\end{table}



\section{Analysis}

We inferred models from the Google implementation of QUIC, specifically version 39. For all experiments, we used the source code from the Chromium repository\footnote{\url{https://github.com/chromium/chromium}} with commit tag \texttt{e611939ed2}. 

In our analysis, we make use of two different input alphabets, resulting in two different models. The first, minimal input alphabet does not contain the 0-RTT message. Thus, it contains all messages from Table~\ref{tbl:messages} except the \texttt{0RTT-CHLO}. This makes model learning easier because the mapper does not need to store the received \texttt{REJ} tags. This eliminates some potential bugs. The second, more extensive alphabet allows models in which a connection can also be started using a previously received \texttt{REJ} message.

In both tests we use the same configuration for StateLearner. We use L* as the learning algorithm. For the equivalence test, we use random queries. The minimum input length is set at 5 and the maximum is 10. The number of test queries that should be made in the equivalence test is set to 100. 

\begin{figure*}
	\centering
	\includegraphics[width=0.8\linewidth]{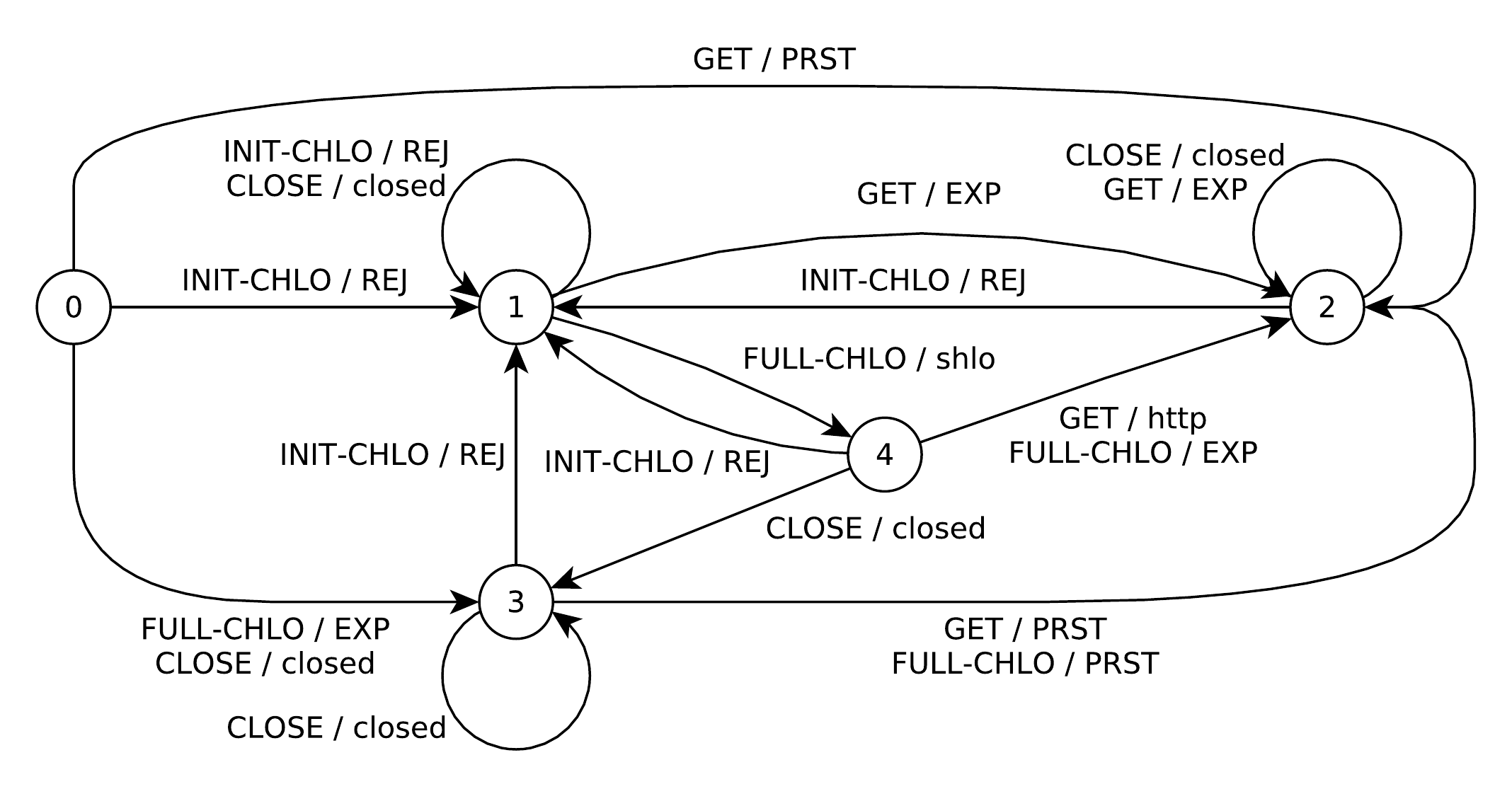}	
	\caption{Inferred model for the minimal input alphabet without 0-RTT}
	\label{fig:nozrttcleanedmodel}
\end{figure*}

The model inferred for the minimal alphabet can be found in Figure~\ref{fig:nozrttcleanedmodel}. It was learned in 42 minutes using 104 queries. The initial state is state 0 and there are four possible transitions. If the client sends an \texttt{INIT-CHLO} packet, the server responds with a \texttt{REJ} message. The other messages will be ignored by the server, either by responding with a public reset (\texttt{PRST}) or no response at all (message has expired, \texttt{EXP}). After receiving the \texttt{REJ} message we get to state 1. In this state, it is possible for the client to set up new connections by sending \texttt{INIT-CHLO} messages with fresh connection IDs, to which the server will respond with a \texttt{REJ} message every time. The client uses the tag/value pairs from the received \texttt{REJ} message and concatenates it to its initial \texttt{INIT-CHLO} to get a \texttt{FULL-CHLO} message. If the tags are correct, the server responds with a \texttt{SHLO} message. This can be seen in the transition from state 1 to state 4. The other transition from state 1 show an erroneous \texttt{HTTP/2 GET} request. However, the connection has not been established yet. Therefore, the server does not respond to this request. The outcome of this request is \texttt{EXP}. We do not see any unexpected states if we compare it with the available documentation.

At this point, the connection between the client and the server has been established and an \texttt{HTTP/2 GET} request can be made. This request results in the \texttt{HTTP} response, as can be seen in the transition from state 4 to state 2. In this last state we observe some curious behavior. We expect that it should be possible to make multiple \texttt{HTTP/2 GET} requests and receive the same \texttt{HTTP} response. However, this seems not to be the case. After making the first \texttt{HTTP} request, the client does not receive any response from the server upon subsequent requests. Manual inspection of the server actions showed that it would respond the second and third time with an acknowledgement to the request. 

As this behavior is not mentioned in the documentation, we tried to find its cause. First, it could be that requesting the same origin on the same stream ID causes the server not to respond. However, it is not possible to change the stream on which the request is made. QUIC uses stream 3 for transmitting compressed headers for all other streams. This helps in processing of the headers~\cite{Britt2016}. Another option was to change the domain name and test whether we could make two consecutive GET requests to different domains. However, this is not possible as the Server Name Indication (SNI) tag binds the connection to a specific domain. We believe that the server expects some client-side caching. Unfortunately, we were not able to verify this.

\begin{figure*}
	\centering
	\includegraphics[width=\linewidth]{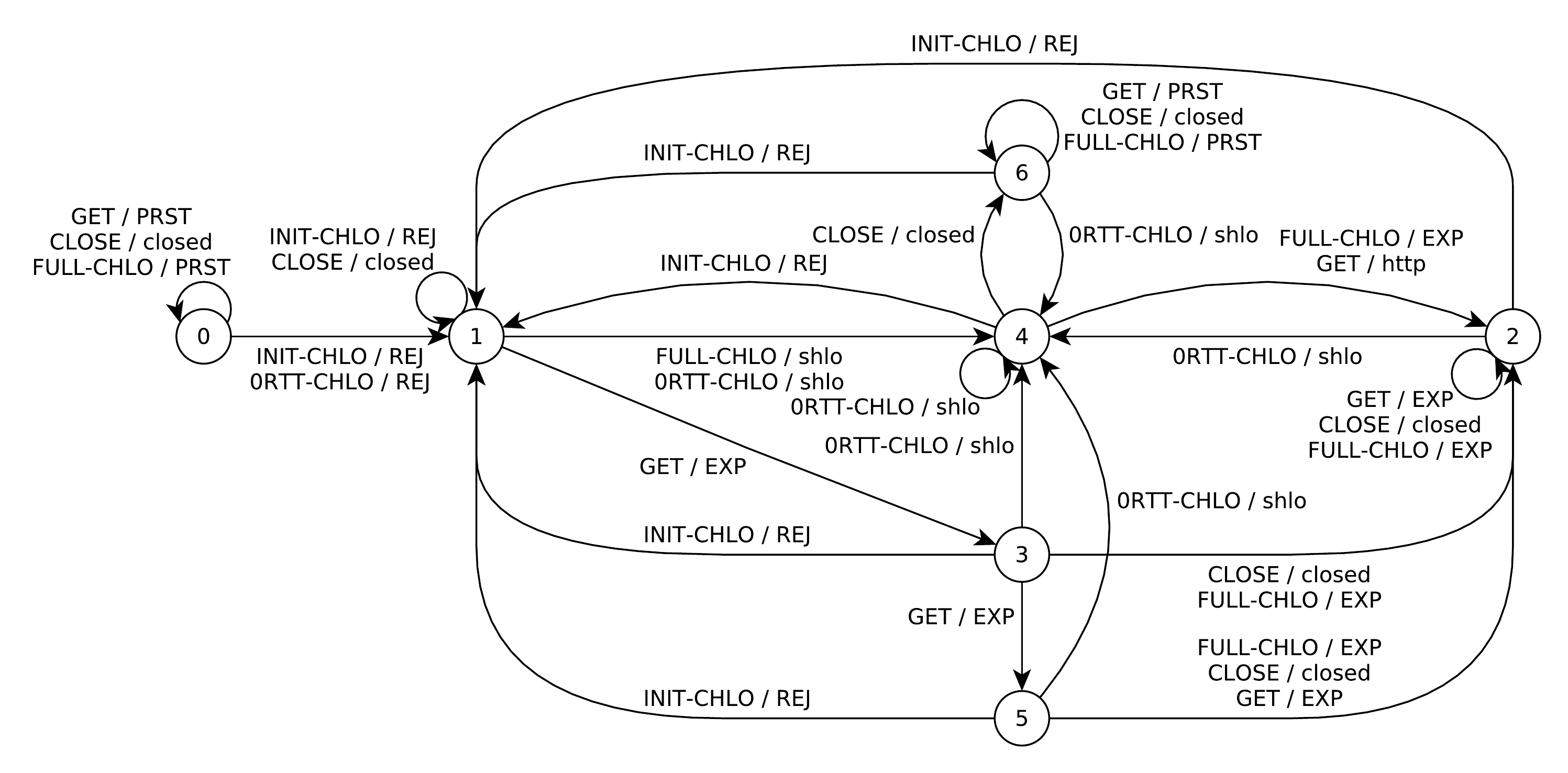}	
	\caption{Inferred model fro the extended input alphabet with 0-RTT}
	\label{fig:cleanedmodel}
\end{figure*}

Next, we look at the model inferred with the extended input alphabet, including the 0-RTT, which can be found in Figure~\ref{fig:cleanedmodel}. It took 73 minutes and 225 queries to learn this model. Compared to the previous model, we see one new input symbol. This is \texttt{0RTT-CHLO}, which represents the message that is the result of the concatenation of the previously received \texttt{REJ} tag/value pairs to an \texttt{INIT-CHLO} request. Once the client has established a connection with the server, it can use this message to set up a connection directly. We can see this in state 4. After sending the complete \texttt{CHLO} request in state 1, the client has knowledge of all the required tags and values by the server. It can send the \texttt{0RTT-CHLO} which results in a \texttt{SHLO} and make a \texttt{HTTP2 GET} request. To achieve true 0-RTT, the client needs to make this request after sending the \texttt{0RTT-CHLO} without waiting for the response of the server. Also in this model, we do not see any unexpected states. 

\sloppypar While learning this model, we encountered some additional non-determinism. A \texttt{0RTT-CHLO} message changes in behavior over the course of three consecutive requests. The first time it triggers a \texttt{REJ} response, the second time it uses the received message to craft a complete \texttt{CHLO} message, which then results in a \texttt{SHLO} message. The third time it also results in a \texttt{SHLO}, because of the new connection ID it chooses. If we use the majority approach to responses, a \texttt{0RTT-CHLO} message always returns a \texttt{SHLO}---but this is not correct. We could not send the \texttt{0RTT-CHLO} three times back; so, we have to find a different approach to achieve response certainty. We perform some manual inspection and apply some response filtering. For example, it is not possible to receive an \texttt{HTTP} response to an initial \texttt{CHLO} message. This response might actually be a retransmission to a previous request. This response would then be discarded and a correct one is returned to the learner. 

\section{Conclusion}

In this paper, we have applied state machine inference to QUIC. To the best of our knowledge, this is the first study to provide such a model specification. This approach works in a completely black-box fashion, making it suitable for any QUIC implementation regardless of implementation details, including the programming language or the operating system. To infer models for the QUIC server, we have handled issues regarding non-deterministic behavior. The resulting state machines are useful for manual analysis, to understand the protocol implementation and to analyze its correctness. Although we have used model learning on a particular, recent QUIC implementation, the resulting state machines could also be included in regression testing. As a result, unintended changes in the implemented state machine will be detected automatically.

Our method offers more insights into the QUIC server's possible states, in particular, those in Google's implementation. In future work, we would like to extend the mapper to also learn the client side of a QUIC implementation. Moreover, as the IETF is working on a standardized version of QUIC, we expect also other parties to implement the protocol soon. It will therefore be possible to analyze and compare new QUIC implementations. 


%
%


\bibliographystyle{plain}
\bibliography{reference}

\begin{thebibliography}{10}

\bibitem{angluin1987learning}
Dana Angluin.
\newblock Learning regular sets from queries and counterexamples.
\newblock {\em Information and computation}, 75(2):87--106, 1987.

\bibitem{carlucci2015http}
Gaetano Carlucci, Luca De~Cicco, and Saverio Mascolo.
\newblock {HTTP} over {UDP}: an experimental investigation of {QUIC}.
\newblock In {\em Proc. of Symposium on Applied Computing (SAC)}, pages
  609--614. ACM, 2015.

\bibitem{Britt2016}
Britt Cyr, Jeremy Dorfman, Ryan Hamilton, Jana Iyengar, Fedor Kouranov, Charles
  Krasic, Jo~Kulik, Adam Langley, Jim Roskind, Robbie Shade, Satyam Shekhar,
  Cherie Shi, Ian Swett, Raman Tenneti, Victor Vasiliev, Antonio Vicente,
  Patrik Westin, Alyssa Wilk, Dale Worley, Fan Yang, Dan Zhang, and Daniel
  Ziegler.
\newblock {QUIC} wire layout specification.
\newblock Technical report, Google, 2016.

\bibitem{deruiter15-tls}
Joeri de~Ruiter and Erik Poll.
\newblock Protocol state fuzzing of {TLS} implementations.
\newblock In {\em Proc. of USENIX Security Symposium}, 2015.

\bibitem{fischlin2014multi}
Marc Fischlin and Felix G{\"u}nther.
\newblock Multi-stage key exchange and the case of {Google's QUIC} protocol.
\newblock In {\em Proc. of SIGSAC Conference on Computer and Communications
  Security (CCS)}, pages 1193--1204. ACM, 2014.

\bibitem{fiteruau16-tcp}
Paul Fiter{\u{a}}u-Bro{\c{s}}tean, Ramon Janssen, and Frits Vaandrager.
\newblock Combining model learning and model checking to analyze {TCP}
  implementations.
\newblock In {\em Proc. of International Conference on Computer Aided
  Verification (CAV)}, pages 454--471. Springer, 2016.

\bibitem{grigorik2013high}
Ilya Grigorik.
\newblock {\em High Performance Browser Networking: What every web developer
  should know about networking and web performance}.
\newblock O'Reilly, 2013.

\bibitem{Kakhki:2017:TLL:3131365.3131368}
Arash~Molavi Kakhki, Samuel Jero, David Choffnes, Cristina Nita-Rotaru, and
  Alan Mislove.
\newblock Taking a long look at {QUIC}: An approach for rigorous evaluation of
  rapidly evolving transport protocols.
\newblock In {\em Proc. of Internet Measurement Conference (IMC)}, pages
  290--303. ACM, 2017.

\bibitem{langley2017quic}
Adam Langley, Alistair Riddoch, Alyssa Wilk, Antonio Vicente, Charles Krasic,
  Dan Zhang, Fan Yang, Fedor Kouranov, Ian Swett, Janardhan Iyengar, et~al.
\newblock The quic transport protocol: Design and internet-scale deployment.
\newblock In {\em Proc. of Special Interest Group on Data Communication
  (SIGCOMM)}, pages 183--196. ACM, 2017.

\bibitem{lychev2015secure}
Robert Lychev, Samuel Jero, Alexandra Boldyreva, and Cristina Nita-Rotaru.
\newblock How secure and quick is {QUIC}? {Provable} security and performance
  analyses.
\newblock In {\em Proc. of Security and Privacy (S\&P)}, pages 214--231. IEEE,
  2015.

\bibitem{esorics2018-wifi}
Chris McMahon~Stone, Tom Chothia, and Joeri de~Ruiter.
\newblock Extending automated protocol state learning for the 802.11 4-way
  handshake.
\newblock In {\em Proc. of European Symposium on Research in Computer Security
  (ESORICS)}, pages 325--345. Springer, 2018.

\bibitem{megyesi2016quick}
P{\'e}ter Megyesi, Zsolt Kr{\"a}mer, and S{\'a}ndor Moln{\'a}r.
\newblock How quick is {QUIC}?
\newblock In {\em Proc. of International Conference on Communications (ICC)},
  pages 1--6. IEEE, 2016.

\bibitem{niese2003integrated}
Oliver Niese.
\newblock {\em An integrated approach to testing complex systems}.
\newblock PhD thesis, Universit{\"a}t Dortmund, 2003.

\end{thebibliography}

\end{document}